# Liquid Crystal Colloids

Ivan I. Smalyukh

*Department of Physics, Department of Electrical, Computer, and Energy Engineering, Soft Materials Research Center and Materials Science and Engineering Program, University of Colorado, Boulder, CO 80309, USA*

*Renewable and Sustainable Energy Institute, National Renewable Energy Laboratory and University of Colorado, Boulder, CO 80309, USA*

*Email: ivan.smalyukh@colorado.edu*

**Abstract**

Colloids are abundant in nature, science and technology, with examples ranging from milk to quantum dots and the "colloidal atom" paradigm. Similarly, liquid crystal ordering is important in contexts ranging from biological membranes to laboratory models of cosmic strings and liquid crystal displays in consumer devices. Some of the most exciting recent developments in both of these soft matter fields emerge at their interface, in the fast-growing research arena of liquid crystal colloids. Mesoscale self-assembly in such systems may lead to artificial materials and structures with emergent physical behavior arising from patterning of molecular order and nano- or micro-particles into precisely controlled configurations. Liquid crystal colloids show an exceptional promise of new discovery that may impinge on the composite material fabrication, low-dimensional topology, photonics, and so on. Starting from physical underpinnings, I review the state of art in this fast-growing field, with a focus on its scientific and technological potential.

## 1. INTRODUCTION

Colloids are soft matter systems in which microscopic particles of one substance are suspended throughout another substance (1). These gas, solid or liquid particles, each with typical dimensions ranging from nanometers to micrometers, are kept suspended in the host surrounding medium by thermal fluctuations (1). The widely familiar naturally occurring colloids include fog, paint, and milk, whereas colloidal dispersions of synthetic particles, such as quantum dots and plasmonic nanoparticles, promise to revolutionize modern technologies and consumer products ranging from the new generations of solar cells to information displays (1-3). Colloids can form liquid, crystalline and glassy condensed matter states (1), which often serve as model systems to understand the physics of their molecular and atomic counterparts. Colloids also offer the possibility of harnessing self-assembly to realize unusual composite materials and to understand complex behavior of biological systems (1-3).

Liquid crystals (LCs), another important class of soft matter systems, are formed by weakly interacting anisotropic building blocks, such as molecules or micelles with anisometric rod-lie or disc-like shapes (1). These molecules self-organize in states with orientational order that are liquids (they flow), but exhibit elastic-type stiffness to certain orientational deformations, thus also resembling anisotropic properties and elastic behavior typically associated with solids (1). Weak external stimuli can dramatically alter interactions between the building blocks of LCs, so that, for example, LCs can melt into isotropic fluids due to modest temperature changes or respond to low-voltage electric fields (1). This facile response to weak electric fields and other external stimuli enables many practical applications of LCs, ranging from information displays to biomedical detectors.

LC colloids are formed by dispersions of colloidal particles in the LC host medium rather than in conventional isotropic liquids (2). The discovery of novel colloidal interactions in LCs (2), which is mediated by the elasticity of the LC medium hosting the colloidal nano- or micrometer-sized particles, prompted the emergence of a new sub-field of soft matter at the interfaces of LC and colloidal sciences. The fast growth of this new research arena is fueled by the richness of fundamental physics phenomena and a strong potential for technological applications. LC colloids further bridge the colloidal and LC sciences with fields as diverse as topology, biology and photonics (2-7). They may enable technological breakthroughs in the development of flexible information displays, efficient conversion of solar energy to electricity, tunable photonic crystals and metamaterials and other novel optically controlled materials capable, in turn, of controlling light, etc. For example, LC colloids may allow for realizing artificial composite materials with pre-engineered optical and mechanical properties defined through an exquisite control of structural organization of metal and semiconductor nanoparticles at the mesoscopic scales, as we discuss later in Section 8 of this review. The excitement that LC colloids bring from the fundamental science standpoint stems from their hierarchy of important length and time scales, leading to creation of entirely new concepts and generalizations. Two decades since its origin (2, 8-15), the field of LC colloids became a well-established fast-growing area of soft matter research. The main goal of this review is to summarize the recent key developments in this field and to overview the dazzling variety of current activities and future developments emerging on the horizon, with a focus on colloidal dispersions with building blocks of the LC much smaller than colloidal particles (**Figure 1***a*).

## 2. LC-COLLOIDAL INTERFACES, BOUNDARY CONDITIONS AND ORIENTATIONAL ELASTICITY

One key difference between the conventional isotropic-host-based and LC colloids is that the molecular interactions at the interfaces of colloidal particles and the surrounding LC are highly anisotropic (**Figure 1**). These interactions are characterized by the easy-axis orientation of the director **n** (average direction of ordering of anisotropic building blocks like molecules in LC phases) at the particle's surface with respect to the local surface normal, which depends on particle's chemical composition and surface morphology and is controlled by chemical functionalization and other means. Similar to the case of flat surfaces (1,15-17), boundary conditions (BCs) on colloidal particles can define energy-minimizing local orientation of **n** to be tangential, perpendicular or tilted with respect to the surfaces (**Figure 1b**). In a uniformly aligned nematic LC, the director field **n**(**r**) is subjected to both the BCs on particle surfaces and the far-field uniform alignment of the director, $\mathbf{n}_0$. BCs are characterized by the surface anchoring energy and a corresponding coefficient *W*. Typically $W<1$ mJ/m$^2$, much smaller than the isotropic part of the corresponding surface energy. For example, in the case of perpendicular BCs with the easy axis **v** (a unit vector) locally orthogonal to the particle surface, the surface anchoring free energy can be expressed as an integral over the particle surface (18)

$$F_{anchoring} = -\frac{W}{2}\int(\mathbf{v}\cdot\mathbf{n})^2 dS. \qquad 1.$$

The bulk free energy due to orientational elasticity of a nematic LC reads (1,18):

$$F_{elastic} = \int\left\{\frac{K_{11}}{2}(\nabla\cdot\mathbf{n})^2 + \frac{K_{22}}{2}[\mathbf{n}\cdot(\nabla\times\mathbf{n})]^2 + \frac{K_{33}}{2}[\mathbf{n}\times(\nabla\times\mathbf{n})]^2\right\}dV, \qquad 2.$$

where $K_{11}$, $K_{22}$ and $K_{33}$ are Frank elastic constants that pertain to splay, twist and bend distortions of **n**(**r**) and the integration is done over the sample's volume. The LC-particle interactions are determined by a competition of bulk elastic and surface anchoring energies, which is characterized by the so-called "extrapolation length" $\xi_e=K/W$, where $K=(K_{11}+K_{22}+K_{33})/3$ is the average elastic constant (typically within 1-20pN, say $K\approx6.5$pN for pentylcyanobiphenyl) and $W$ is typically within $10^{-4}$–$10^{-6}$ J/m$^2$. When the colloidal particle's diameter $D<<\xi_e$ ($\xi_e$ is within 50-10,000 nm), it induces no defects and distortions of $\mathbf{n}(\mathbf{r})$ can be neglected (**Figure 1c**). The director in this case stays aligned along $\mathbf{n}_0$ while violating BCs on the colloidal surface. In contrast, elastic distortions and bulk or surface topological defects occur when $D>>\xi_e$, in which case the BCs are met and matched to $\mathbf{n}_0$ by distorting $\mathbf{n}(\mathbf{r})$ around the particle. The emergence of line and point defects in the latter case (1,12,18) is illustrated using the example of perpendicular BCs in **Figure 1d,e**. Overall, the structural configurations of $\mathbf{n}(\mathbf{r})$ are determined by minimization of the total free energy, the sum of Equations 1 and 2. Since the bulk elastic free energy roughly scales as $\propto KD$ and the surface energy as $\propto WD^2$, the total free energy minimization leads to violations of BCs for small $D$ and $W$ and to elastic distortions and defects in the opposite case. Similar considerations also apply to the cases of tangential and tilted BCs. The nature of particle-induced defects depends on particle's topology and geometry, as discussed in Section 7.

## 3. ELASTIC MULTIPOLES

Multipole expansion, which physicists and engineers learn as part of an undergraduate-level introduction to electromagnetism (19), is widely used in many branches of physics and cosmology. In describing the far-field distortions in electric, magnetic, gravitational and other fields, say due to localized charge or current distributions, truncation of the multipole expansion to its leading-order non-zero term is useful for theoretical calculations and explaining physical behavior (19). The leading-order multipoles are typically monopoles, dipoles, and quadrupoles, with higher-order multipoles rarely playing dominant roles, though we will see that they can be important for certain types of nematic colloids (2,8-13,20-24). Much like in the case of electrostatic charge distributions, the far-field distortions of $\mathbf{n}(\mathbf{r})$ due to a colloidal particle can be represented as elastic multipoles (2,13,20-24). Far from a colloidal particle, the deviations of the director $n_\mu$ ($\mu=x; y$) from the far-field uniform director $\mathbf{n}_0=(0, 0, 1)$ are small (24,25). Assuming one-elastic-constant approximation and representing the field as $\mathbf{n}(\mathbf{r}) \approx (n_x, n_y, 1)$, the LC elastic free energy is expressed as

$$F_{elastic} = \frac{K}{2} \sum_{\mu=x,y} \int d^3\mathbf{r} \nabla n_\mu \cdot \nabla n_\mu . \qquad 3.$$

The corresponding Euler-Lagrange equations arising from the minimization of $F_{elastic}$ are of the Laplace type, $\Delta n_\mu = 0$, with solutions expanded into multipoles:

$$n_\mu(\mathbf{r}) = \sum_{l=1}^{N} a_l (-1)^l \partial_\mu \partial_z^{l-1} \frac{1}{r}, \qquad 4.$$

where $a_l = b_l r_0^{l+1}$ is the elastic multipole moment of the $l$-th order ($2^l$-pole), $r_0$ is particle's radius and expansion coefficients $b_l$ can be found from experiments (24,25). This electrostatics analogy provides a framework for understanding, predicting, and engineering both the director structure and self-assembly of nematic colloids.

At no external fields, the lowest order elastic multipoles are dipoles (2,13), such as the ones formed by a colloidal microsphere with perpendicular BCs and a bulk hedgehog point defect (**Figures 1e** and **2a-c**). This particle-induced dipolar structure of $\mathbf{n}(\mathbf{r})$ (**Figure 2a-c**) is invariant with respect to rotations around $\mathbf{n}_0$ but lacks mirror symmetry with respect to a plane passing through the particle's equator orthogonally to $\mathbf{n}_0$. Microspheres with perpendicular BCs can also induce a disclination loop defect around the particle's equator, often called "Saturn ring", which yields an elastic quadrupole (**Figure 2*d-f***). In contrast to the elastic dipoles, this quadrupolar **n**(**r**)-structure (**Figure 2*e***) is mirror-symmetric with respect to the plane passing through the particle's equator orthogonally to **n**$_0$. Colloidal spheres with strong tangential anchoring at their surface also form **n**(**r**)-distortions of quadrupolar type (**Figure 2*g-i***). They induce two boojums at the poles along **n**$_0$. For weak surface anchoring BCs or particle diameter comparable to $\xi_e$, both the tangential and perpendicular BCs lead to quadrupolar configurations, similar to the elastic quadrupole with the "Saturn ring" or boojums, but the defects become "virtual" (25,26) (within particle's volume), as the director is allowed to deviate away from the easy axis orientations. Particles with tilted conically degenerate BCs (**Figure 2*j-l***) locally distort **n**(**r**), so that the polarizing optical micrographs (**Figure 2*j***) feature eight bright lobes around the particle, separated by eight dark regions within which **n**(**r**) at the particle's perimeter is parallel to **n**$_0$. These features of polarizing micrographs are different from the cases of dipoles and quadrupoles that exhibit alternation of two bright and two dark regions (**Figure 2*a***) and four bright and four dark regions (**Figure 2*d,g***), respectively. The tilt of **n**(**r**) away from **n**$_0$ switches between clockwise and counterclockwise directions eight times as one circumnavigates the hexadecapole (**Figure 2*l***), which is different from dipoles, where it switches two times, and quadrupoles, where it switches four times. Both the surface boojums at the particle's poles along **n**$_0$ as well as the Saturn ring at the particle's equator are induced by hexadecapoles (24). To minimize the free energy cost of bulk elastic distortions, interaction of conically degenerate BCs on the microsphere with the uniform **n**$_0$ lifts the conical degeneracy of surface anchoring and yields an axially symmetric **n**(**r**) that can be thought of as a superposition of the quadrupolar structures with boojums and Saturn ring (**Figure 2*l***) (24).

To illustrate the analogies between elastic and electrostatic multipoles, a projection $n_x$ of

**n(r)** onto the *x*-axis orthogonal to $\mathbf{n}_0$ can be visualized using colors that highlight positive, near-zero and negative $n_x$ (**Figure 2c,f,i,k**). Away from the particle surface and singularities such as boojums, hedgehogs and disclination loops, **n(r)** is continuous (24). The maps of $n_x$, plotted on spherical surfaces encompassing particles and defects, clearly illustrate the multipolar nature of elastic distortions similar to that in the electrostatics analogues (19). The color presentations of elastic distortions induced by colloidal dipoles (**Figure 2c**), two different quadrupoles (**Figure 2f,i**), and hexadecapoles (**Figure 2k**) resemble the corresponding dipolar, quadrupolar, and hexadecapolar electrostatic charge distributions described by $\sigma_l^m(\theta,\phi) = A\cos(m\phi)P_l^m(\cos\theta)$ with $(l, m) = (1, 1), (2, \pm1)$ and $(4, 1)$, respectively, where $-l \leq m \leq l$, $A$ is a normalization constant, $\theta$ is a polar and $\phi$ is an azimuthal angles, and $P_l^m(\cos\theta)$ is the associated Legendre polynomial. Similar to electrostatic charge distributions, the odd moments vanish when **n(r)** is symmetric about the particle center, as in the cases of elastic quadrupoles and hexadecapoles presented in **Figure 2e,h,l** (which have coefficients $b_1$, $b_3$ and $b_5$ and so on equal zero, as required by symmetry). However, both odd and even moments are present for particles with asymmetric **n(r)**, such as elastic dipoles, for which all $b_l$ can be nonzero (13, 21-24). The design of elastic colloidal multipoles not only takes advantage of symmetry considerations, but also builds on the versatile means of controlling **n(r)** using shape and BCs, so that the coefficients $b_l$ corresponding to lower-order multipoles can be tuned to be close to zero, yielding higher order multipoles. For example, the hexadecapole shown in **Figure 2j-l** was designed by tuning $b_2$ to be close to zero while precluding the dipole and octupole moments on the basis of symmetry discussed above (24). In a similar way, an elastic octupole could be potentially obtained by tuning $b_1$ and $b_2$ to zero while maximizing $b_3$, though this has not been achieved so far. Although it is widely assumed that elastic monopoles cannot be realized without external torques applied to colloidal particles (18), certain particles can effectively behave as elastic monopoles when attached to surfaces and exerting torques on the LC director through surface anchoring (23). Both elastic monopoles and octupoles require further studies.

## 4. COLLOIDAL INTERACTIONS MEDIATED BY ORIENTATIONAL ELASTICITY OF LCS

The analogy between electrostatic and elastic multipoles may help devising approaches for self-assembly of mesostructured colloidal composite materials. The potential of elastic interaction between nematic colloidal multipoles, which arises from the changes of free energy (which the medium tends to minimize) due to superposition of elastic distortions induced by interacting particles, is also analogous to that between electrostatic multipoles (2,13,21,24). For example, the potential energy of interaction between two nematic colloidal hexadecapoles scales as $P_8(cos\theta)/R^9$ (24), that between two quadrupoles as $P_4(cos\theta)/R^5$ (8-13,20-22), two octupoles as $P_6(cos\theta)/R^7$ (24,25) and two dipoles as $P_2(cos\theta)/R^3$ (2,13), where $P_{2l}(cos\theta)$ are Legendre polynomials and $R$ is the center-to-center separation distance. Mixed interactions between elastic multipoles of different order can be also predicted by exploiting the electrostatic analogy and, for example, scale as $P_6(cos\theta)/R^7$ between quadrupoles and hexadecapoles. Experimentally, colloidal interactions can be probed using a combination of laser tweezers and video microscopy, as shown for the case of hexadecapoles in **Figure 3**. Interaction between two nematic colloidal hexadecapoles (24) is predicted to have 8 angular sectors of attraction and 8 sectors of repulsion as the center-to-center separation vector **R** is rotated (**Figure 3a,b**) by $2\pi$ with respect to $\mathbf{n}_0$, which is indeed observed by releasing particles from laser traps and directly observing the directionality of interaction forces with video microscopy (**Figure 3c**). Tracking particle motion as they interact reveals that the experimental time dependencies of *R* and distance dependencies of interaction potential (**Figure 3***d*) at different orientations of **R** are all consistent with the model (24). Similar characterization of pair interactions was also done for elastic dipoles and quadrupoles, leaving no doubts that multipolar expansion can provide the necessary foundation for understanding physics of nematic colloids (2, 12, 27-33). However, only some of the elastic multipoles are widely studied, with the most common ones shown in **Figure 2**, and the others often require particles with nonspherical shapes or patterned surface BCs; for example, elastic dipoles with moments orthogonal to $\mathbf{n}_0$ could be induced by triangular and pentagonal platelets but so far could not be realized using microspheres (28). The description of structure and interactions between nematic colloidal particles using Legendre polynomials $P_l^m$ or spherical harmonics with indices (*l*,*m*), where $-l \leq m \leq l,$ also resembles the description of atomic orbitals of elements in the periodic table (**Figure 3***e*), where the analogous indices (*l*,*m*) are quantum numbers. Therefore, the multipolar analysis of nematic colloids not only brings about many interesting analogies and helps to assess the diversity of structures that can be realized, but is also of interest from the standpoint of the colloidal atom paradigm (3-5), as discussed next.

## 5. NEMATIC "COLLOIDAL ATOM" PARADIGM AND COLLOIDAL CRYSTALS

Since the works of Einstein and Perrin showed how particles in colloidal dispersions obey the same statistical thermodynamics as atoms, the colloid-atom analogy has provided insights into physics of atomic and molecular systems through probing dynamics of colloidal crystals and glasses (5). This colloidal atom paradigm has inspired the development of colloidal self-assembly to reproduce or exceed the diversity of atomic systems (3, 33-35). LC colloids have a series of advantages in addressing these grand scientific and engineering challenges. The LC host naturally allows for realizing anisotropic inter-particle interactions (2), for spontaneously aligning shape-anisotropic particles on the basis of surface anchoring interactions (28,33), and for exploiting rich multipolar behavior of elastic interaction forces (2), which would be hard or impossible to achieve in conventional isotropic fluids (3). Furthermore, arrays of topological defect lines within the ground-state cholesteric blue phases, or in other LC phases when stabilized by confinement, can be used to entangle or entrap colloidal particles to form various sparse crystals, quasicrystals and other assemblies (34, 36-41). In conventional colloidal dispersions, high-symmetry crystals with 3D cubic and 2D hexagonal lattices have been studied extensively (3-7, 42), confirming and extending Einstein's colloidal-atom analogy by giving direct experimental confirmations of theoretical models originally developed for atomic and molecular crystals (3,5). However, experimental realization of colloidal architectures with low crystallographic symmetry remained challenging, even though colloids are capable of designed control of their self-assembly through DNA functionalization (35). A series of self-assembled 2D and tetragonal 3D crystal lattices and other colloidal architectures have been achieved in LC colloidal crystals using "guided" assembly with laser tweezers (42-48), though these colloidal superstructures could be only built "particle-by-particle" and only up to several primitive cells in dimensions because of relying on laser tweezers rather than the self-assembly alone. However, it was recently shown that low-symmetry triclinic colloidal crystals can self-assemble in LC dispersions as a result of competition between the highly anisotropic elastic interactions discussed above and (more conventional) electrostatic repulsive forces (33,43). Realization of this triclinic colloidal crystal suggests that the large crystallographic diversity previously established for atomic and molecular crystals can now be accessible in LC colloids as well.

The triclinic colloidal crystal serves as an illustration of how competition of long-range electrostatic and anisotropic quadrupolar elastic interactions in nematic colloids can lead to self-assembly of colloidal particles with long-range orientational and positional ordering (**Figure 4**) (33). The long-range electrostatic repulsions arise from weakly screened Coulomb-type forces in a low-ionic-strength nonpolar LC with a large (500 nm and larger) Debye screening length. Since the elasticity-mediated interactions are also long-ranged, colloidal crystal assemblies with micrometer-range lattices emerging from the competition of these electrostatic and elastic forces can be an order of magnitude larger than the size of constituent colloidal particles (**Figure 4a-d**) (33). The crystallographic axes of the triclinic lattices and colloidal nanorods within them tend to follow the director, which can be controlled on large scales via approaches used to manufacture LC displays (18). Triclinic crystallization of particles at packing factors <<1% shows a potential for fabricating mesostructured composites through self-assembly on device scales, tuned by weak external stimuli, such as low voltages (43). From a fundamental science standpoint, it is interesting that the thermal fluctuations of particles within the lattice and the melting and crystallization transitions (**Figure 4e,f**) are consistent with what is expected for their atomic counterparts (33). Mean square displacement (MSD) exhibited by colloidal particles within these dispersions saturates with time when in crystal but continues growing when in a colloidal fluid state (inset of **Figure 4f**). Characterization of MSD $\langle \Delta r^2(t) \rangle$ relative to the crystal lattice, quantified by the dependence of the Lindemann parameter $\delta L=[3\langle \Delta r^2(t\rightarrow\infty)\rangle/(4R^2)]^{1/2}$ on the nanorod number density $\rho_N$, provides a means of demonstrating how low-symmetry positional ordering (**Figure 4f**) emerges from the interplay of weakly screened electrostatic and elasticity-mediated interactions in a nematic host. The triclinic pinacoidal lattices of orientationally ordered nanorods (**Figure 4**) are the lowest symmetry 3D colloidal crystals achieved so far (**Figure 4a-e**) and may prompt realization of LC colloidal architectures with other symmetries, ranging from cubic to triclinic pedial lattices.

Though demonstrated only for elastic quadrupoles and electrostatic monopoles so far, the use of competition of elastic, electrostatic and other (e.g. magnetic) interactions to guide symmetry and density of colloidal crystals can be extended to other nematic colloidal multipoles, like dipoles and hexadecapoles (**Figure 2**), which were previously exploited mostly only from a standpoint of interactions guided by laser tweezers (24,44-48). Considering the intrinsic anisotropy and diversity of elastic multipoles, one can expect a large variety of low-symmetry colloidal architectures emerging from such interactions. Furthermore, because atomic orbitals and the outmost occupied electron shells of chemical elements (**Figure 3***e*) are described by spherical harmonics $Y_{lm}(\theta,\phi)$, they are analogous (in terms of symmetry) to the nematic "colloidal atoms." Since the symmetry of the "director wiggle wave function" of nematic hexadecapolar colloids (**Figure 2***k*,*l*) would correspond to chemical elements with filled *g*-shells (**Figure 3***e*), which have not been discovered so far, nematic colloidal atoms already exceed the diversity of their atomic counterparts (24).

## 6. MOLECULAR-COLLOIDAL ANISOTROPIC FLUIDS

Realization of thermodynamic phases that combine low-symmetry order and fluidity is one of grand challenges in the soft matter research. Both molecular and conventional colloidal LCs are classic examples of how various degrees of orientational and partial positional order can be combined with fluidity in nematic, smectic, and columnar phases (18), though they tend to be nonpolar and exhibit high symmetry. Starting from the early works of de Gennes and Brochard (49), researchers predicted and demonstrated self-alignment of shape-anisotropic colloidal particles like rods and platelets in LCs (28,33,36,50-56). This LC-mediated spontaneous orientation of anisotropic particles relative to $\mathbf{n}_0$ emerges from elastic (49,55) and/or surface-anchoring-based (56) interactions. It allows the system minimizing surface anchoring energy at the LC-colloidal interface given by Equation 1 or elastic energy due to the bulk distortions nearby the particle induced BCs at its surface given by Equation 2, respectively, though surface anchoring and elastic energies are often comparable and both contribute to defining particle orientations (57-68). In most cases, dilute dispersions of anisotropic colloidal particles in LCs do not alter the symmetry of the LC dispersion as compared to that of the LC as the orientations of particles typically simply mimic the ordering of the LC host (69) and can be treated in terms of the behavior of individual inclusions (70-88). LC-colloidal interactions, along with the direct magnetic inter-particle interactions, can also lead to the polar alignment of magnetic colloidal inclusions (49-54), although the orientations of the magnetic dipoles of colloidal particles are commonly slaved to $\mathbf{n}$, orienting either parallel or perpendicular to it without breaking uniaxial symmetry. Experimental realization of low-symmetry ordered fluids, like orthorhombic biaxial LCs (89), is challenging for both molecular and colloidal systems, especially in the case of nematic phases without positional ordering. However, similar to the case of low-symmetry colloidal crystals (33), LC colloidal dispersions of various nanoparticles also provide a series of advantages and promise realizing hybrid LC-colloidal fluids with low-symmetry orientational order (16). For example, a recently introduced molecular-colloidal complex fluid formed by a dispersion of magnetic nanoplates in a thermotropic nematic LC exhibits co-existing polar and biaxial ordering of organic molecular and magnetic colloidal building blocks, with the lowest $C_s$ symmetry orientational order in a fluid medium, which so far could not be realized in purely molecular or colloidal systems alike (16).

In the molecular-colloidal LC with the $C_s$ symmetry, guided by interactions at different length scales (16), rod-like organic molecules of the host fluid spontaneously orient along the director while magnetic colloidal nanoplates order with their dipole moments **m** (**Figure 5***a-d*) parallel to each other, but pointing at an angle to the director, yielding macroscopic magnetization at no external fields (**Figure 5***e,f*). This unusual physical behavior is enabled by the control of conically degenerate surface anchoring at the magnetic nanoparticle surfaces (16,17), which assures tilted orientations of magnetic moments **m** and magnetization **M** with respect to $\mathbf{n}_0$. Magnetic hysteresis loops (**Figure 5***e,f*) and facile switching of such ordered ferromagnetic fluids emerge from competing actions of elastic and magnetic torques and rich magnetic domain (16). This and other molecular-colloidal complex LC fluids promise rich behavior arising from the properties of solid nanoparticles and their long-range ordering prompted by interactions with a host medium at the mesoscale. Future research may lead to hybrid molecular-colloidal soft matter phases with different point group symmetries describing order of molecular and colloidal building blocks, such as an orthorhombic biaxial (89) nematic molecular-colloidal fluid in which anisotropic nanoparticles (like nanorods) could orientationally order in a direction orthogonal to the ordering direction of rod-like LC molecules. Such hybrid molecular-colloidal systems, in which colloidal self-organization is enriched by elastic and surface anchoring interactions, are expected to yield soft condensed matter phases that could not be realized previously.

## 7. TOPOLOGY, GEOMETRY AND CHIRALITY: SURFACES WITH AND WITHOUT BOUNDARY

The interaction of colloidal particles with the LC director field can be understood and generalized on the basis of the interplay between the topologies of fields and surfaces widely studied by mathematicians (90). All surfaces are characterized (up to homeomorphism) by their genus, orientability, and number of boundary components, as stated by the classification theorem (90). Colloidal particles with all dimensions larger than $\xi_e=K/W$ produce well defined BCs and can be considered having orientable surfaces without boundaries, so that their interaction with the **n(r)** depends on the particle surface's genus $g$ (or, equivalently, its Euler characteristic $\chi=2-2g$). Colloidal particles tend to have surfaces with spherical or other topologically simple shapes with $g=0$ and $\chi=2$, such as prisms (continuously deformable to spheres). However, one can fabricate colloidal particles with different numbers of handles and $g\neq0$ (91-95). Fabrication approaches include conventional photolithography (28), direct laser writing (91,92), two-photon polymerization (93,94) and laser-induced reduction of colloidal graphene oxide nanoflakes (95). In LCs, these particles induce 3D **n(r)** and topological defects dictated by colloidal topology. 3D nonlinear optical imaging and numerical modeling (91-94) reveal that topological charge induced by the particles is conserved and that the net hedgehog charge $m$ of particle-induced bulk defects obeys predictions of the Gauss-Bonnet and Poincaré-Hopf index theorems. Although particles can induce different **n(r)**-structures, depending on confinement, external fields and material properties, the sum of hedgehog charges due to induced point defects and disclination loops, $\sum_i m_i = -m_c = \pm\chi/2$, always compensates the colloidal particles' hedgehog charge $m_c$ due to **n(r)** at their surfaces and is uniquely pre-determined by particle's topology (**Figure 6**). The use of "±" signs in this relation depend on a choice of the vectorizing **n(r)**-field direction, but the relative signs of the induced defects are fixed. This relation allowed for establishing a procedure for the assignment and summation of topological charges in 3D director fields (91).

Particles with tangential BCs additionally induce boojums at the LC colloidal interface.

The total winding number of boojums in the projection of **n(r)** to the particle surface, **n$_s$(r)**, is always $\Sigma_i s_i = \chi$, also satisfying topological theorems (92). These simple relations describe defects induced both by the *g*=0 spherical surfaces (**Figures 1** and **2**) and by surfaces like handlebodies and torus knots with *g*≠0 (**Figure 6**), though additional self-compensating defects can appear to reduce the energetic cost of elastic distortions associated with matching BCs on colloidal surfaces to **n$_0$**. For example, $\Sigma_i s_i = 0$ for six self-compensating boojums induced by a trefoil knot colloidal particle (**Figure 6*g-i***), as expected for the torus knot's surface with $\chi$=0. However, beyond the predictions of topological theorems, it was found that nematic colloids can also have mutually tangled physical knots of particles and line defects in **n(r)**, as illustrated in **Figure 6*j-l*** (93). For all nematic colloidal particles, the interplay of topologies of surfaces, fields, and defects guides **n(r)** to comply with the particle shape, generating knotted, linked, and other 3D configurations (91-94), providing experimental insights into the aspects of the mathematical knot theory (90,93,94).

Nematic colloidal particles in the form of thin foil with two lateral dimensions >>$\xi_e$ but one dimension (thickness) <<$\xi_e$ behave as analogs of orientable surfaces with boundaries, to which the constrains on induced defects discussed above cannot apply and, generally, no defects need to be induced (91,96). Yet, by morphing shapes of such foil-based particles, one can pattern the defects to achieve various elastic multipoles and drive LC-mediated self-assembly similar to that exhibited by nematic colloidal particles with surfaces without boundaries (96). They behave differently from both the microparticles (larger than $\xi_e$) with strong BCs (Figs. 1d,e, 2 and 3) and nanoparticles (smaller than $\xi_e$) with effectively weak surface anchoring (**Figures 4** and **5**), being capable of inducing no defects or various self-compensating defect structures, depending on their geometric shape and orientation relative to **n$_0$** (96). Non-orientable colloidal particles with boundary have also been recently theoretically considered (97), though their experimental realization still presents a challenge.

Colloidal particle's shape plays a key role in defining the locations and types of induced defects even when topology of particle surfaces stays intact (**Figure 7**). For example, by shaping $g=0$ colloidal particles with perpendicular BCs as truncated pyramids (**Figure 7a-d**), one pre-selects the topology-complying defect with $m=\pm 1$ to take form of a disclination loop localized near the pyramid's base, as well as uniquely defines orientation of the ensuing elastic dipole (**Figure 7d**). Polygonal platelets (**Figure 7e-j**) with odd and even numbers of edge faces behave as dipoles and quadrupoles, respectively, illustrating that the elastic multipole nature can be controlled via tuning the particle's geometric shape (28). Particles shaped as letters of the Latin alphabet (**Figure 7k,l**), on the other hand, exhibit well-defined orientations with respect to $\mathbf{n_0}$ and induce surface boojums with $\Sigma_i s_i = \chi$ correlated with their shapes. Other colloidal shapes studied recently include pyramids, spirals, shape-morphing elastomeric rods, gourd-shaped dimers, particles with nanoscale surface roughness, and so on (98-113), in all cases demonstrating that the geometric shape pre-defines locations of topological defects and colloidal interactions.

Often closely related to particle shape and topology, chirality of nematic colloids can also play an important role in defining their physical behavior. For example, knot-shaped particles shown in **Figure 6g-l** are chiral in nature and induce 3D director distortions that mimic this chirality (93,94), similar to surface-attached chiral particles studied in Ref. (23). However, even mirror-symmetric particles can induce 3D chiral distortions of $\mathbf{n(r)}$ due to nontrivial conformation states and transformations of the topological defects (102). Chiral nature of LC hosts also drives formation of chiral distortions in $\mathbf{n(r)}$ and chiral colloidal superstructures. For example, it can lead to formation of twist-escaped disclination loops around particles (114), formation of sparse assemblies enabled by cholesteric quasi-layers (40), topological solitons bound to particles (115,116), and other types of chirality-enabled colloidal behavior (117-126).

## 8. PROPERTY ENGINEERING IN NEMATIC COLLOIDAL COMPOSITES

Self-organization of nano-sized functional units is an exceptionally promising way of designing artificial composite materials with new macroscopic physical behavior (127). Many approaches focus on the material behavior arising from ordering, co-localization and alignment of nanoparticles that lead to controlled plasmon-exciton and other interactions enabled by the nanoscale confinement (127). In the field of LC colloids, recent studies explore such interactions at the levels of individual quantum dot, upconversion and plasmonic nanoparticles (128-138) and the novel composite materials (43,55,56,61,63,64,66,67,69,73,80,139-144), demonstrating a means of controlling nanoparticle and composite optical characteristics, such as luminescence lifetimes. For example, plasmon-exciton interactions can alter blinking statistics of quantum dot nanoparticles and can shorten the luminescence lifetimes by an order of magnitude (128). The host medium's intrinsic order translates into the self-organization of immersed inclusions depending on shapes/sizes of particles and how anisotropic colloidal structures interact with the molecular alignment fields, as discussed above. Shape-dependent colloidal interactions and ordering of quantum dots, dye-doped nanostructures and plasmonic metal nanoparticles reveal underpinning physical mechanisms that can ultimately determine composite properties (55,61). Facile control of nanoparticle and molecular organization and the ensuing composite properties can be achieved by applying electric, magnetic, and optical fields (80). In addition to the equilibrium structures and various crystal or LC colloidal phases, future designs of composite material behavior may also involve various types of LC colloidal dynamics (145-150), though additional studies are still needed to make this possible.

## 9. CONCLUSIONS AND OUTLOOK

Mimicking and ultimately exceeding diversity of organization of atomic elements is one of the main goals of colloidal science, which could allow for self-assembly of a broad range of mesostructured composite materials. Self-organization of colloidal nano- and micrometer-sized particles dispersed in LCs is a promising approach to achieve this goal. Due to unusual but controlled molecular interactions at colloidal surfaces, the solid particles can locally perturb the uniform molecular alignment of the nematic host, forming elastic multipolar distortions of ordering that drive highly anisotropic interactions and lead to low-symmetry organization in the form of colloidal crystals with tunable lattice parameters and colloidal fluids with orientational ordering. The interdisciplinary research on LC colloids addresses some of the most exciting problems at the interfaces of materials engineering, soft matter physics, nanoscience, and photonics. The emerging scientific frontiers at the nexus of these fields show exceptional promise of significant new discovery and applications. They will advance our fundamental knowledge of self-assembly in complex fluids and will impinge on fields as diverse as information displays, metamaterial and photonic crystal fabrication and energy conversion.

**DISCLOSURE STATEMENT**

The author is not aware of any affiliations, memberships, funding, or financial holdings that might be perceived as affecting the objectivity of this review.

**ACKNOWLEDGMENTS**

I gratefully acknowledge research contributions and discussions over many years with my former and current students and postdocs working on LC colloids, including P. Ackerman, M. Campbell, J. Evans, D. Gardner, J. Giller, A. Hess, L. Jiang, C. Lapointe, T. Lee, Q. Liu, A. Martinez, H. Mundoor, M. Pandey, S. Park, O. Puls, B. Senyuk, R. Trivedi, O. Trushkevych, C. Twombly, M. Varney, Y. Yuan, and Q. Zhang. I also acknowledge support from the NSF grant DMR-1410735.

**ACRONYMS AND DEFINITIONS LIST**

**Colloids:** microscopically dispersed gas, solid or liquid particles of one substance, each with dimensions ranging from nanometers to micrometers, that are kept suspended throughout another substance by thermal fluctuations

**Liquid crystals (LCs):** soft matter systems that stand between crystalline solids and isotropic fluids and combine the properties of fluidity with orientational ordering

**Nematic phase:** LC phase formed by anisotropic (e.g. rod-like or disc-like) molecules, particles or micelles exhibiting the long-range orientational order but no positional order

**LC director n**: average direction of ordering of anisotropic building blocks of LC phases with nonpolar $\mathbf{n} \equiv -\mathbf{n}$ symmetry

**Surface anchoring:** anisotropic part of the surface energy describing the dependence of surface energy on orientation of **n** at LC surfaces

**Cholesteric phase:** LC fluid locally similar to the nematic phase, but with **n** continuously rotating around an axis called "helical axis"

**LC defect (singularity):** a discontinuity in **n** that cannot be removed via smooth deformations

**Boojum:** a localized point-like singularity at the surface of LC

**Boojum winding number:** the number of times that the director $\mathbf{n}_s$ at the LC-particle interface rotates by $2\pi$ as one circumnavigates the surface boojum's defect core once

**Hedgehog defect:** a localized point-like singularity in **n** in the bulk of LC

**Hedgehog charge** of a region of space $V$ bounded by a surface $S=\partial V$ is the degree of **n** along $S$, calculated by integrating the Jacobian of $\mathbf{n}(\mathbf{r})$ over $S$ as $m = (1/4\pi) \int_S dx_1 dx_2 \mathbf{n} \cdot \partial_1 \mathbf{n} \times \partial_2 \mathbf{n}$

**Disclination:** a discontinuity in the orientational order in the form of a line defect along which **n** cannot be defined

**LC topological soliton:** a nonsingular, continuous, but topologically nontrivial configuration of **n**(**r**) that cannot be transformed to a uniform state through smooth deformations of the field

**Mean square displacement (MSD):** time-average squared displacement of a particle, a measure of the deviation of its position from some reference position

**Lindemann criterion:** melting of crystals occurs because of vibrational instability, when the amplitude of thermal vibrations of building blocks is high compared to distances between them

**Magnetic hysteresis:** a characteristic of ferromagnets associated with the lack of retraceability of the magnetization curve when the magnetic field is relaxed

**Genus *g*:** a maximum integer number of cuttings along non-intersecting closed simple curves without rendering the resultant manifold disconnected, equal to the number of handles on it

**Euler Characteristic $\chi$:** a topological invariant $\chi = 2 - 2g$ that describes a topological space's shape or structure regardless of the way it is bent

**Photon upconversion:** a process of sequential absorption of two or more photons that leads to light emission at a wavelength shorter than that of excitation

**Trefoil knot:** the simplest nontrivial knot obtained by joining together the two loose ends of a common overhand knot, yielding a knotted loop

**Handlebody:** a submanifold of Euclidean space comprising a ball with handles attached to it along its boundary

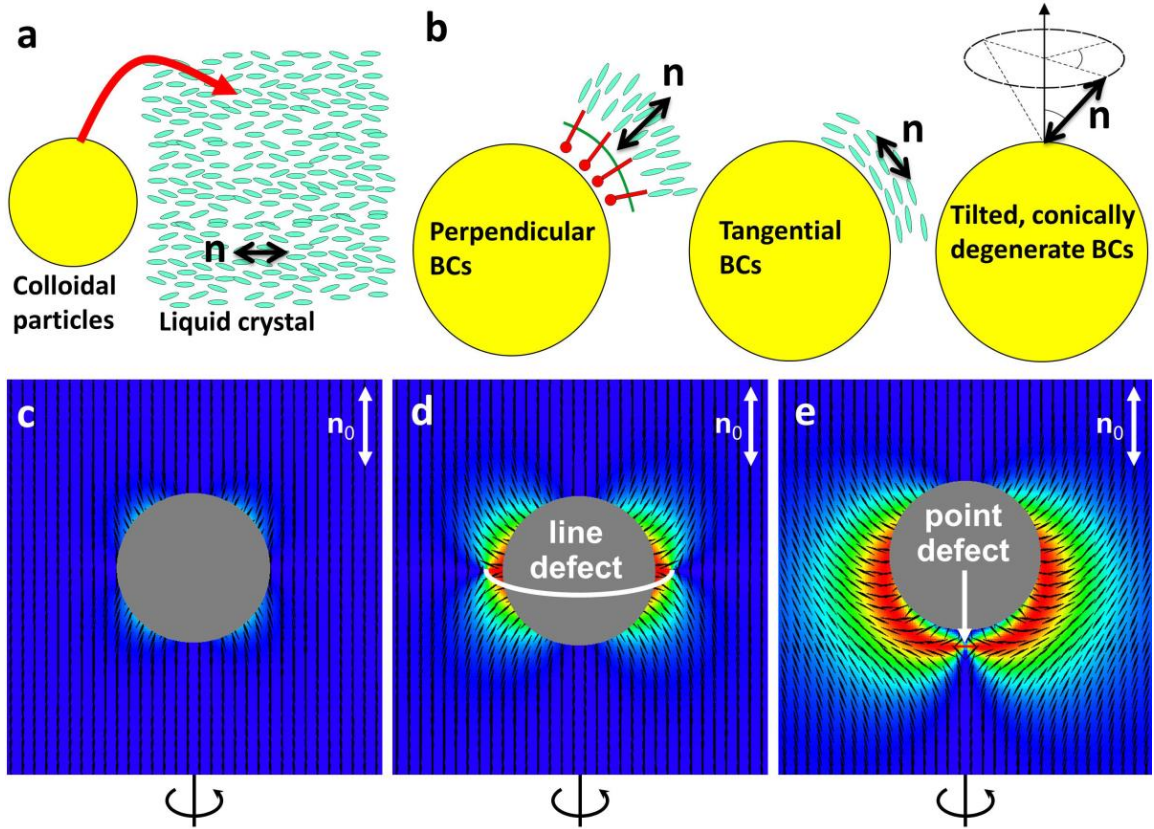

**Figure 1.** LC colloids and control of surface BCs. (*a*) Schematic showing colloidal particles with dimensions roughly in the range 10 nm - 10 μm dispersed in LC composed of rod-like molecules of about 1nm in length. (*b*) Surface functionalization or natural material behavior of particles enables the control of surface BCs for **n** to be (from left to right) perpendicular, tangential, or tilted conically degenerate. (*c-e*) Interplay of elastic and surface anchoring forces defines **n(r)**-configurations, ranging from (*c*) director structure with weak deviations of **n(r)** away from **n**$_0$ and violation of BCs at the particle surface for $D<<\xi_e$ to the structures with (*d*) ring-like line and (*e*) point-like topological defects observed at $D>\xi_e$. The colors highlight deviations of **n(r)** away from **n**$_0$, depicting regions (blue) with **n(r)** along **n**$_0$ and (cyan, green, yellow and red) **n(r)** tilted away from **n**$_0$ to progressively larger angles.

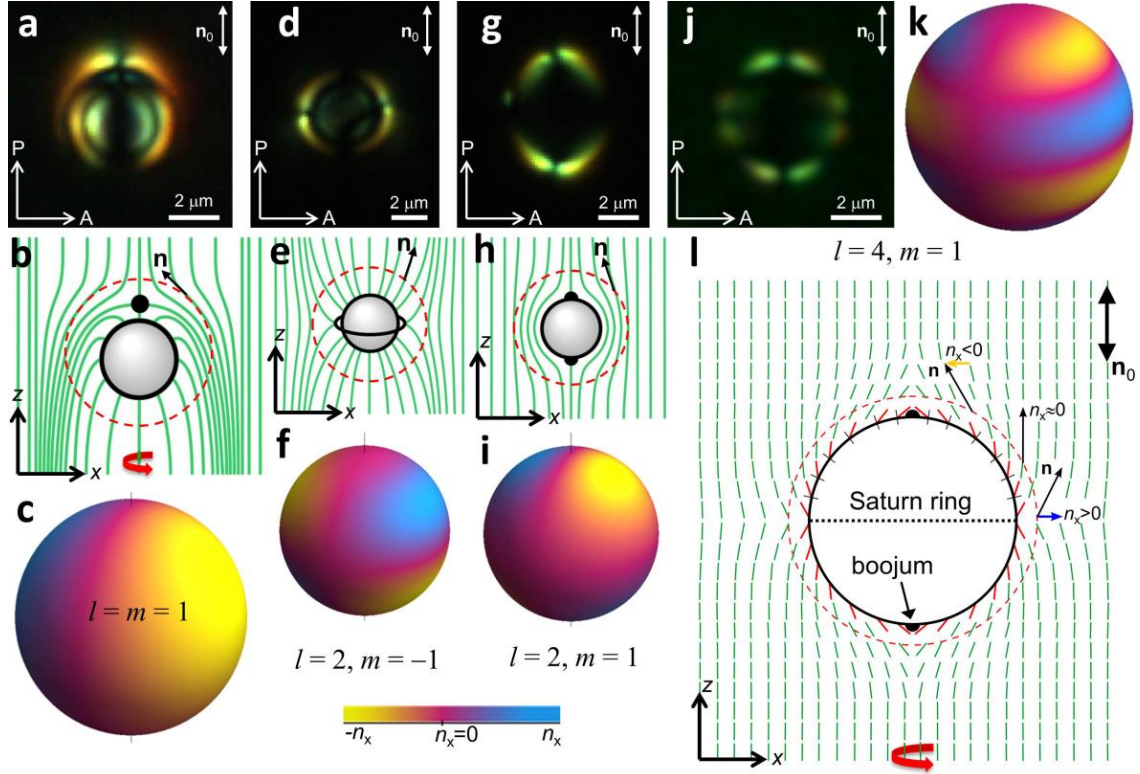

**Figure 2.** Elastic colloidal multipoles. (*a-c*) Elastic dipoles formed by microspheres with perpendicular BCs shown with the help of (*a*) a polarizing optical micrograph obtained between crossed polarizer P and analyzer A, (*b*) axially symmetric director configuration visualized with solid green lines (black filled circle depicts a hedgehog defect), and (*c*) a color-coded map of **n** on a sphere marked with a red dashed circle in (b) and corresponding to a dipole with $l=m=1$. (*d-f*) An elastic colloidal quadrupole formed by a microsphere with perpendicular BCs shown with the help of (*d*) a polarizing optical micrograph, (*e*) axially symmetric **n**(**r**) visualized with green lines (the black ring depicts a disclination), and (*f*) a color-coded map of **n** on a surrounding sphere corresponding to a quadrupole with $l=2$, $m=-1$. (*g-i*) An elastic colloidal quadrupole formed by a microsphere with tangential BCs shown with the help of (*g*) a polarizing optical micrograph, (*h*) axially symmetric **n**(**r**) visualized by green lines (the black hemispheres depict boojums), and (*i*) a color-coded map of **n** corresponding to a quadrupole with $l=2$, $m=1$. (*j-l*) An elastic colloidal hexadecapole formed by a microsphere with tilted conically degenerate BCs shown with the help of (*j*) a polarizing optical micrograph, (*k*) a color-coded map of **n** corresponding to a multipole with $l=4$, $m=1$, and (*l*) an axially symmetric **n**(**r**) depicted with green rods (dashed black line at the equator depicts a disclination loop and the solid black hemispheres depict boojums). For more details, see Ref. (24).

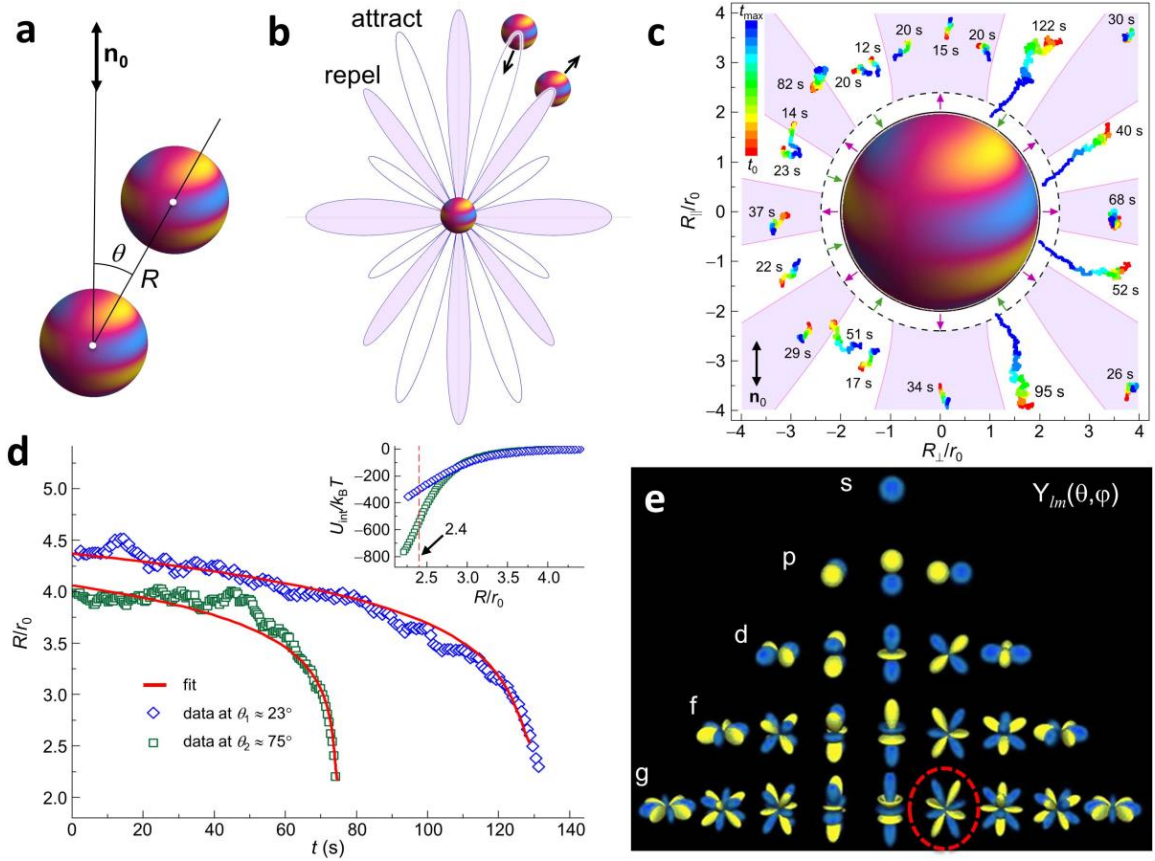

**Figure 3.** Colloidal interactions and analogy with chemical elements. (*a*) Definition of the angle $\theta$ between **R** connecting two colloidal hexadecapoles and $\mathbf{n}_0$. (*b*) A theoretical prediction for the angular sectors of attraction (white) and repulsion (magenta) between a pair of hexadecapoles. (*c*) Experimental test of interactions indeed exhibits eight angular sectors of attraction and eight sectors of repulsion. When released from laser tweezers, hexadecapoles moves towards or away from one another depending on $\theta$, as shown by color-coding time within the particle motion trajectories (top-left inset). (*d*) Time dependencies $R(t)$ for the initial $\theta$ within the attractive sectors and the dependence of inter-particle pair potential $U_{int}$ on separation R normalized by particle radius $r_0$ (inset) are consistent with the theoretical predictions. For more details, see Ref. (24). (*e*) Electron probability distributions characterizing the atomic orbitals described in terms of spherical harmonics $Y_{lm}$ with the corresponding indices as quantum numbers; the *g*-shell configuration with $l=4$, $m=1$ of still undiscovered chemical elements that would have such shells filled is marked using a red dashed circle and is analogous to a nematic colloidal hexadecapole (**Figure 2***j*-*l*).

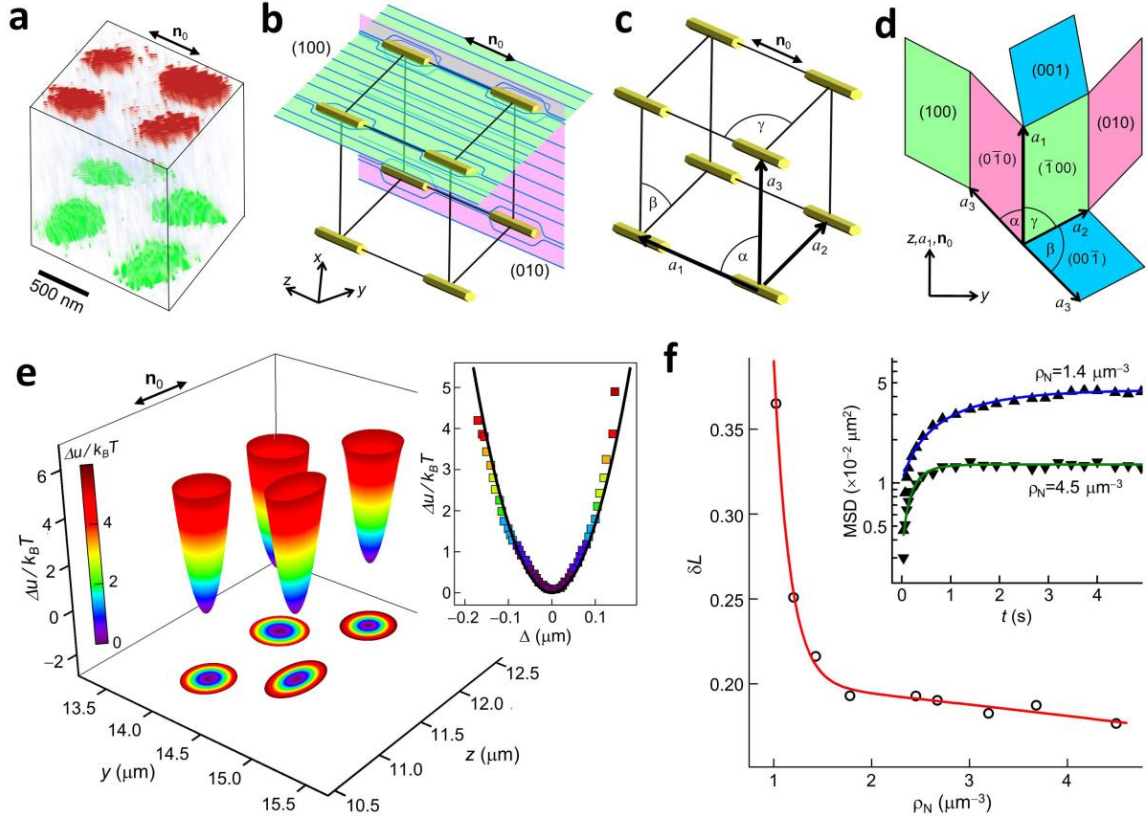

**Figure 4.** Colloidal crystals. (*a*) 3D micrograph showing a primitive unit cell of a triclinic colloidal crystal, which was reconstructed on the basis of confocal imaging. It shows spatial arrangements of colloidal particles as they explore potential energy landscape near their minimum-energy triclinic lattice sites. (*b-d*) Schematics (not to scale) of a primitive cell of a triclinic colloidal crystal, (*b*) showing **n**(**r**)-distortions (blue lines) induced by nanorods, (*c*) defining parameters of a triclinic lattice, and (*d*) showing it unfolded. Crystallographic planes, axes, and angles are marked on the schematics. (*e*) The potential energy $\Delta u$ landscape corresponding to four lattice sites in the (100) crystallographic plane of a triclinic crystal. The inset shows a local distance $\Delta$ dependence of the relative $\Delta u$ experienced by colloidal particles. (f) Lindemann parameter $\delta L$ versus $\rho_N$ of nanorods, characterizing the crystallization-melting transition. Inset shows the MSD of particles versus time at $\rho_N$=1.4 µm$^{-3}$ (▲) and 4.5 µm$^{-3}$ (▼). For more details, see Ref. (33).

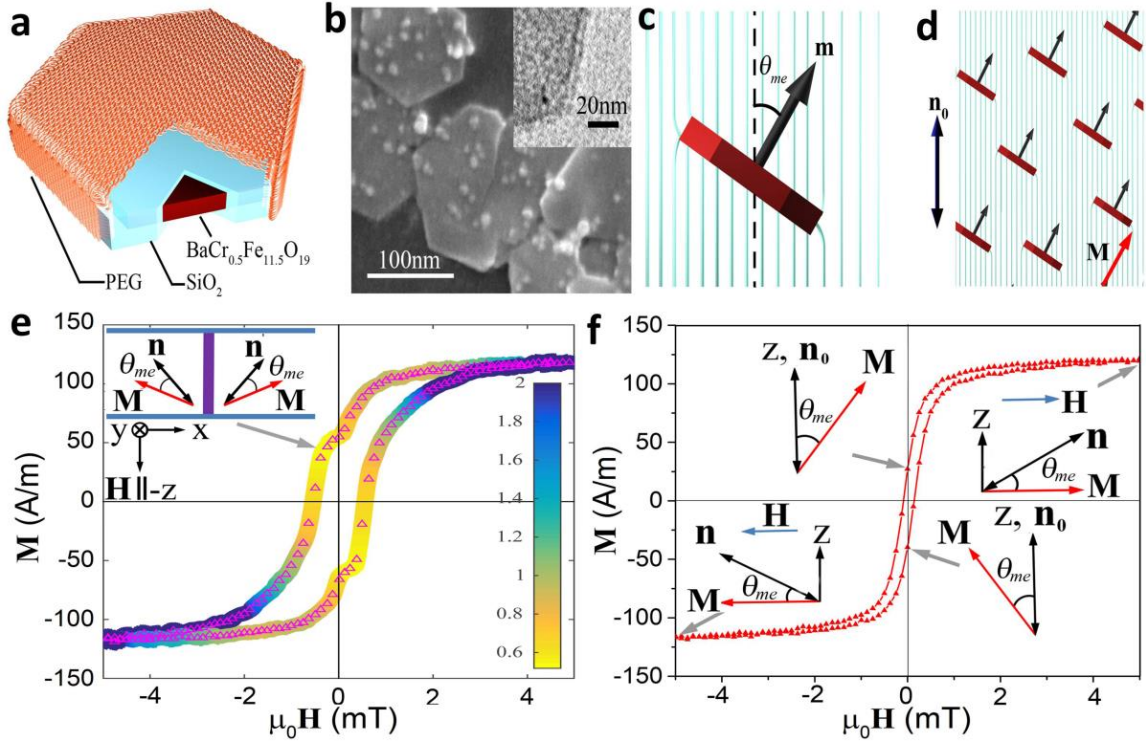

**Figure 5.** Molecular-colloidal LC fluids. (*a*) Schematic of a ferromagnetic nanoplate, coated with silica and then further functionalized with polyethylene glycol (PEG). (*b*) Scanning TEM image of such magnetic nanoplates, with a zoomed-in TEM image of a nanoplatelet's edge revealing the ≈5 nm-thick silica shell (inset). (*c*) Schematic of an individual nanoplate in LC, with the magnetic moment **m** tilted away from **n₀** to an angle $\theta_{me}$. (*d*) A schematic of the biaxial ferromagnetic molecular-colloidal hybrid LC with magnetization **M** tilted away from **n₀**. (*e*) Experimental (open triangle symbols) and theoretical (solid curve, with the relative domain size coded according to the color scheme shown in the right-side inset, where the average lateral size of domains to cell thickness ratio varies from 0.5 to 2) magnetic hysteresis loop measured along **n₀**. Schematics in the inset show relative orientations of **M** and **n** within domains that correspond to the part of hysteresis loop indicated using a gray arrow. (*f*) Hysteresis loop probed for the same sample as in (*e*), but for **H**⊥**n₀**; relative orientations of **M** and **n** corresponding to different parts of the hysteresis loop are indicated using gray arrows. For more details, see Ref. (16).

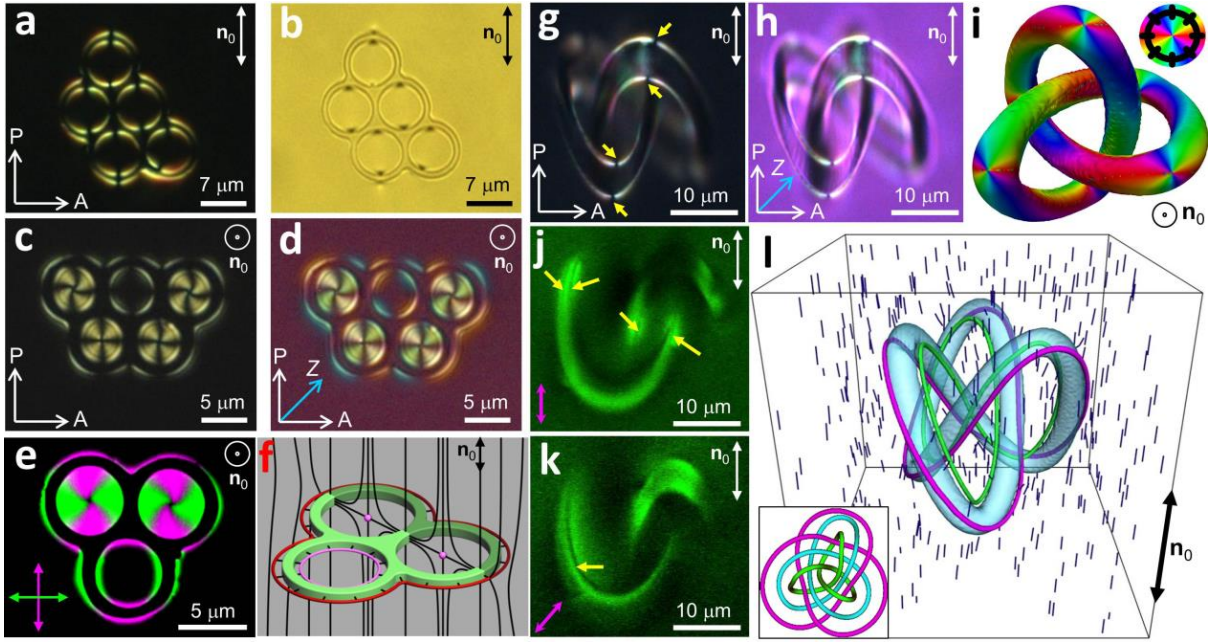

**Figure 6.** Topology-controlled nematic colloidal behavior. (*a,b*) Transmission-mode optical micrographs of a *g*=5 colloidal handlebody with tangential BCs in a nematic LC obtained (*a*) between crossed polarizers and (*b*) without polarizers. Boojums are visible as discontinuities in bright brushes in (*a*) and as dark light-scattering localized regions in (*b*). (*c,d*) Polarizing optical micrographs of a *g*=5 colloidal handlebody with perpendicular BCs in a nematic LC obtained (*c*) between crossed P and A without and (*d*) with an additional 530 nm phase retardation plate. (*e,f*) A *g*=3 colloidal handlebody with perpendicular BCs in a nematic LC (*e*) imaged using three photon excitation fluorescence polarizing microscopy and (*f*) shown schematically. The overlaid fluorescence patterns obtained for the two mutually orthogonal polarizations (green and magenta double arrows) of excitation light are shown using green and magenta colors. Black lines represent **n**(**r**) and the red or magenta points and rings depict topological defects with opposite hedgehog charges. (*g-i*) A trefoil knot particle with tangential BCs in an aligned LC imaged (*g,h*) between crossed P and A (*g*) without and (*h*) with an additional 530 nm retardation plate having slow axis "Z" marked by a blue arrow. Boojums are indicated by yellow arrows in (*g*). (*i*) 3D representation of **n**(**r**) deviating away from **n₀** due to the incorporated trefoil-knot particle. Colors depict the azimuthal orientation of **n**(**r**) when projected onto a plane orthogonal to **n₀** according to the scheme shown in the inset. The structure is visualized on a tube following the knot particle's surface. Points where different colors meet are boojums. (*j-l*) A colloidal trefoil knot with perpendicular BCs. (*j,k*) Three-photon excitation fluorescence polarizing microscopy images of **n**(**r**) for excitation light polarizations (magenta double arrows) at different orientations with respect to **n₀**. Yellow arrows mark the defect lines. (*l*) Computer-simulated **n**(**r**) around a trefoil knot with perpendicular BCs and the torus plane oriented orthogonally to **n₀**. Green and magenta lines show regions with reduced scalar order parameter corresponding to cores of two knotted defect lines (*j,k*). Inset shows a schematic of mutual linking between the particle knot (blue) and defect knots (green and magenta). For more details, see Refs. (92-94).

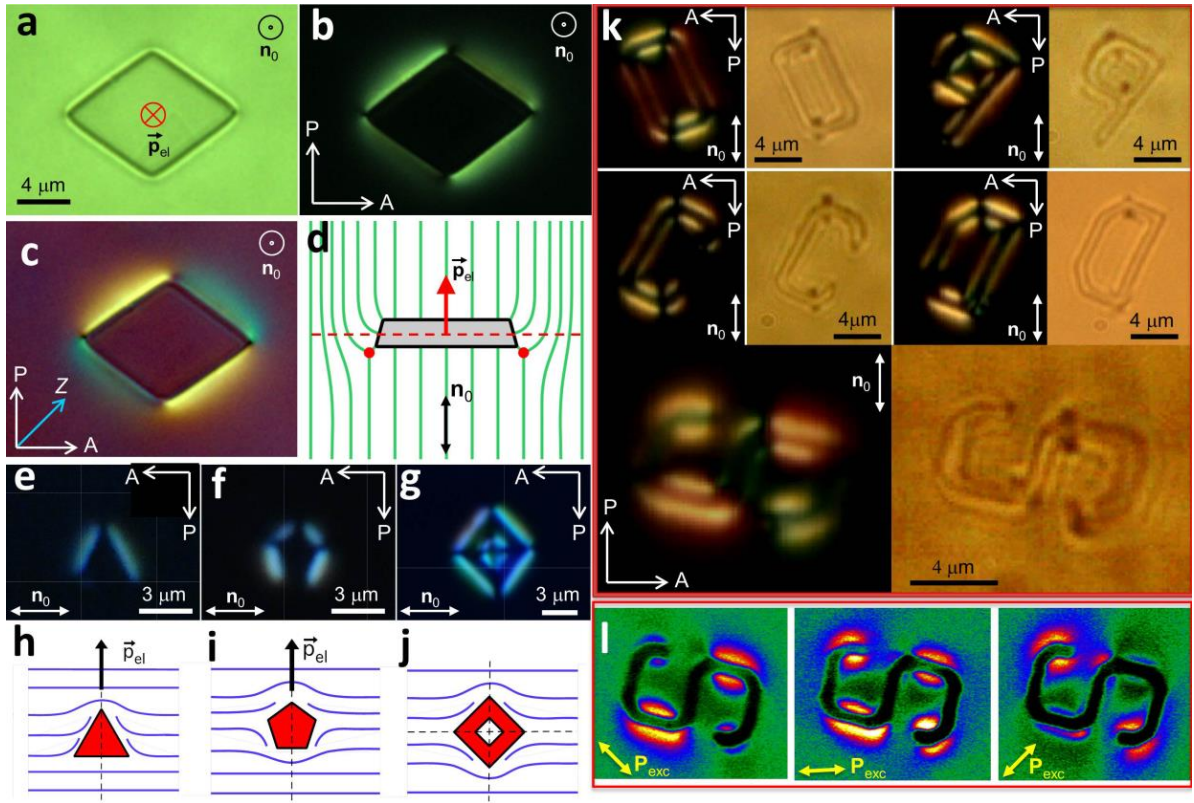

**Figure 7.** Shape-controlled nematic colloidal behavior. (*a-c*) A truncated pyramid with a rhomb-like base imaged using optical microscopy (*a*) without polarizers, (*b*) between crossed polarizers, and (*c*) with a 530nm phase retardation plate. (*d*) **n**(**r**) (green lines) around the particle in a cross-section plane containing $\mathbf{n}_0$; red filled circles are the intersections of a disclination loop with the plane. Red arrow shows the elastic dipole $\mathbf{p}_{el}$. (*e-g*) Polarizing optical micrographs of **n**(**r**) induced by polygonal prisms with tangential BCs and (*e*) three, (*f*) five, and (*g*) four edge faces. (*h-j*) Schematic illustrations of the respective **n**(**r**) structures (blue lines) around the prisms that correspond to (*h,i*) elastic dipoles for odd numbers of edge faces and (*j*) to elastic quadrupoles for even numbers of edge faces. (*k*) Optical micrographs of particles with shapes of Latin alphabet letters immersed in an aligned LC and viewed (left-side images) between crossed P and A and (right-side images) without polarizers; boojums are visible as discontinuities in bright brushes in the former and as dark localized regions in the latter. (*l*) Three photon excitation fluorescence polarizing microscopy images of **n**(**r**) around the "S"-shaped particles obtained for linear polarizations of excitation light $\mathbf{P}_{exc}$ (yellow double arrows). For more details, see Refs. (28,36,73).